%% file: malware-eprint.tex
\documentclass[10pt,journal,cspaper,compsoc]{IEEEtran}

\usepackage{times,url,algorithm,algorithmic,graphicx,amsmath}

\ifCLASSOPTIONcompsoc
\else
\fi
\ifCLASSINFOpdf
\else
\fi
\begin{document}
%
\title{Tap-Wave-Rub: Lightweight Malware Prevention for Smartphones using Intuitive Human Gestures\thanks{Tap, Wave and Rub is a magical trick \cite{twr-magic}. In this paper, these ``magical'' gestures are shown to provide an effective defense to the prevailing problem of mobile phone malware.}}
%
%
%
%

\author{Haoyu Li, Di~Ma,~\IEEEmembership{Member,~IEEE,}
        ~Nitesh~Saxena,~\IEEEmembership{Member,~IEEE}, Babins~Shrestha, and Yan~Zhu
\IEEEcompsocitemizethanks{\IEEEcompsocthanksitem Haoyu Li, Di Ma, and Yan Zhu are with the College of Engineering and Computer Science, University of Michigan-Dearborn, Dearborn, MI 48128.\protect\\
E-mail: \{haoyul,dmadma,yanzhu\}@umd.umich.edu
\IEEEcompsocthanksitem Nitesh Saxena and Babins Shrestha are with University of Alabama at Birmingham, Birmingham, AL 35294.\protect\\
E-mail: saxena@cis.uab.edu}
\thanks{The work of Haoyu Li, Di Ma, and Yan Zhu was partially supported by the grant from US National Science Foundation (NSF CNS-1153573).}}

%
%

\markboth{Cryptology ePrint Archive}%
{Shell \MakeLowercase{\textit{et al.}}: Bare Demo of IEEEtran.cls for Computer Society Journals}
%


\IEEEcompsoctitleabstractindextext{%
\begin{abstract}
Malware is a burgeoning threat for smartphones. It can surreptitiously access
sensitive services on a phone without the user's consent, thus compromising the
security and privacy of the user. The problem is exacerbated especially in the
context of emerging payment applications, such as NFC services. Traditional
defenses to malware, however, are not suitable for smartphones due to their
resource intensive nature. This necessitates the design of novel mechanisms
that can take into account the specifics of the smartphone malware and 
smartphones themselves.

In this paper, we introduce a lightweight permission enforcement approach --
\textit{Tap-Wave-Rub (TWR)} -- for smartphone malware prevention. TWR is based
on simple human gestures that are very quick and intuitive but less likely to
be exhibited in users' daily activities. Presence or absence of such gestures,
prior to accessing an application, can effectively inform the OS whether the
access request is benign or malicious.  Specifically, we present the design of
two mechanisms: (1) \textit{accelerometer-based phone tapping detection}; and
(2) \textit{proximity sensor based finger tapping, rubbing or hand waving
detection}. The first mechanism is geared for NFC applications, which usually
require the user to tap her phone with another device. The second mechanism
involves very simple gestures, i.e., tapping or rubbing a finger near the top
of phone's screen or waving a hand close to the phone, and broadly appeals to
many applications (e.g., SMS). In addition, we present the \textit{TWR-enhanced Android
permission model}, the prototypes implementing the underlying gesture
recognition mechanisms, and a variety of novel experiments to evaluate these
mechanisms. Our results suggest the proposed approach could be very effective
for malware detection / prevention, with quite low false positives and false
negatives, while imposing little to no additional burden on the users.  

\end{abstract}

}

\maketitle

\IEEEdisplaynotcompsoctitleabstractindextext

%
\IEEEpeerreviewmaketitle

\section{Introduction} \label{sec:intro}
\input{intro}

\section{Related Work} \label{sec:related}
\input{background}

\section{Background and Models} \label{sec:ourapproach}
\input{threat}
\input{design}
\input{overview}

\input{systemmodel}

\section{Tap-Wave-Rub Gesture Detection} \label{sec:triggering}
\input{tapping}
\input{waving}

\section{Implementation and Evaluation} \label{sec:experiment}
\input{setup}

\input{tap-experiment}

\input{wave-experiment}


\section{Discussion} \label{sec:discuss}
\input{discussion}

\section{Conclusion and Future Work} \label{sec:conclusion}

In this paper, we introduced a lightweight permission enforcement approach -- Tap-Wave-Rub (TWR) -- for smartphone malware detection and prevention. TWR is based on simple
human gestures that are very intuitive but less likely to be exhibited in users' daily activities. Presence or absence of such gestures, prior to accessing a service, can effectively inform the OS whether the access request is benign or malicious.  Specifically, we presented the design of two mechanisms: (1) phone tapping detection based on accelerometer data, and (2) finger tapping, rubbing or hand waving detection based on proximity sensor data. The first
mechanism is geared for NFC applications, which usually require the user to tap her phone with another device. The second mechanism involves very simple gestures from the user, i.e., tapping or rubbing a finger near the top of
phone's screen or waving a hand close to the phone, and broadly appeals to many applications (e.g., SMS). In addition, we present the TWR Android permission model, the
prototypes implementing the underlying gesture recognition mechanisms, and a variety of novel experiments to evaluate them. 

Our results suggest the proposed approach could be very effective for malware prevention, with quite low false positives and false negatives, while imposing little to no additional burden on the users. The false negatives are expected
to further reduce significantly as users become more familiar with the
underlying gestures, especially since they are quite intuitive. In addition, the
false positives can also be carefully avoided in most cases, for example, by
detecting the orientation of the device.


Our future effort will be focused on realizing this approach
in practice and further evaluate it with a wide range of smartphones and smartphone users. On a broader perspective, we believe that the gestures introduced in this paper will also facilitate novel ways in which users can interact with their devices and open up new opportunities (for security
purposes or otherwise).

\bibliographystyle{abbrv}

{\small{
\bibliography{all,malware}
}







\end{document}

%% file: intro.tex
%
Smartphones are undoubtedly becoming ubiquitous. They are not only used as
(traditional) mobile phones for phone calling and SMS messaging, but also for
many of the same purposes as desktop computers, such as web browsing, social
networking, online shopping and banking. Also, smartphones are incorporating
more and more sensors and communication interfaces. Such new capabilities
enable smartphones with many new unique functionalities that desktop computers
lack. For example, many smartphones are beginning to incorporate Near Field
Communication (NFC) chips \cite{nfc}, which allows short, paired transactions
with other NFC-enabled devices in close proximity. The use of NFC-equipped
smartphones as payment tokens (such as Google Wallet) is considered to be the
next generation payment system and the latest buzz in the financial industry
\cite{mobilepayment}. 


Due to their popularity, smartphones are becoming a burgeoning target for
malicious activities. There has been a rapid increase in mobile phone malware
targeting different smartphone platforms
\cite{cabir,commwarrior,viver,soundcomber,touchlogger,OwHnDsPrZh2012,HnOwNgPrZh2012,spiphone}.
After infecting a phone, malware can gain access to sensitive resources /
services on the phone for various purposes, such as stealing user data, making
premium phone calls or SMS, damaging the device, tracking user's activities or
location, or simply annoying the user. 

Newer functionalities of smartphones
only make them more attractive to malware writers. For example, the
incorporation of NFC chips on smartphones provides malware authors another
(possibly much easier) way to deploy their attacks through the NFC interface
\cite{mulliner2009vulnerability}. Especially, due to the ease with which
financial transactions can take place using NFC, it is predicted that NFC will
become a popular target for malware aiming at credential and credit card theft
\cite{felt2011survey}.  Indeed, a proof-of-concept Trojan Horse electronic
pickpocket program under the cover of a tic-tac-toe game has already been
developed by Identity Stronghold \cite{epickpocket}. In this attack, the game
containing the malware is downloaded and installed on a NFC-enabled smartphone.
Once activated (when the game is played), the malware accesses the NFC chip and
enables the RFID (Radio Frequency ID) reading functionality. This reader then
surreptitiously scans tags (e.g. RF tagged credit card) around it and reports
the acquired information to the malware owner through e-mail once a victim tag
is found in proximity. 

While the security community has been battling with PC malware for many
years, malware detection on smartphones turns out to be an even more
challenging problem \cite{burguera2011crowdroid}. This is partially due to the
resource constraints of smartphones (especially limited battery power). Thus,
existing malware defenses for desktop computers cannot be applied directly on
the smartphone platform. Much of the existing research focuses on optimizing desktop based defenses for mobile phones
\cite{venugopal2006efficient,schmidt2009static,Shamili2010malware,cheng2007smartsiren,burguera2011crowdroid}.
These approaches either try to speed-up the detection process using advanced data structures and algorithms, or use the network or remote server to reduce
computation overhead on individual devices. 

In practice, to protect mobile phones from malware attacks, major mobile phone
manufacturers, such as Google, Apple, and Nokia, employ permission models
to prevent malware from being installed at the first place. 
However, this approach relies upon user diligence and awareness, while most computer users lack these traits in practice. 
Instead of relying on user permissions, smartphone manufacturers also rely upon
application review before releasing to people for download. However,
application review process can be cumbersome and prone to human error
\cite{burguera2011crowdroid}.

\subsection{Motivation and Rationale} \label{sec:motivation}
We argue that existing malware defenses, without considering the special
characteristics of smartphone malware and that of smartphones themselves, might
not be sufficient to detect sophisticated malware, such as the pickpocket
malware targeting NFC mentioned previously.\footnote{Throughout the paper, we
will center our malware mitigation design based on properties observed from the
pickpocket malware \cite{epickpocket}; however, our approaches, being
fundamental in nature, will be applicable to a broad range of future malware.}
First, the pickpocket malware \cite{epickpocket}, under the cover of
tic-tac-toe, is quite stealthy.  Its surreptitious scanning may not cause
substantial changes (such as sharp increase in the number of emails sent or in
power consumption) to the normal behavioral profile and therefore behavioral
detection schemes will not be effective. Moreover, most existing malware
detection schemes employ a \textit{posteriori approach}. That is, malicious
attacks are detected after they take place as traces need to be collected and
trained before they can be compared with profiles to find abnormalities.
Because of the sensitive (financial) nature of the NFC service, it is quite
risky to adopt such a \textit{posteriori} detection approach. Instead, it is
important to develop a preventive approach which can constantly monitor,
identify, and then stop such potentially malicious activity \textit{before} it
is launched so as to minimize damage or loss. 


This motivates us to design a novel approach for malware prevention through
contextual awareness. Our rationale is as follows.
%
Smartphones are personal devices. That is, the end user is a human being. Thus,
(legitimate) access to sensitive/valuable services such as premium calls, SMS
or NFC usually involves different types of human activities such as dialing a
phone number, typing a message, or clicking an application icon on the screen
(to execute the application). In contrast, one common pattern followed by
malware found on mobile phones is that it attempts to access sensitive services
without the user's awareness and authorization (thus user activity is very
unlikely to be involved). 
Therefore, one way to detect such unauthorized, thus
potentially malicious behavior, is to validate \emph{whether an action is
initiated by pure software or purposefully by a human user}. 









%

Since legitimate access to sensitive services usually involves different types
of hand movements, we explore the use of gestures to differentiate between pure
software and human-initiated activities. 
In particular, in this paper, we propose \textit{Tap-Wave-Rub (TWR)}, a
lightweight malware detection mechanism for smartphones based on intuitive
human gesture recognition, using sensors already available on current
smartphones with little or no additional user involvement.


The proposed gesture-based detection mechanism serves as an extension to the
currently adopted permission model used by major smartphone OSs. That is,
whenever a sensitive service is requested, a particular gesture needs to be
detected (to make sure it is a human generated activity) before the request can
be granted.  Otherwise, the activity is very likely generated by malware.  As
gesture detection is enforced every time a sensitive request is received, the
proposed mechanism provides continuous monitoring of sensitive resources and
services from unauthorized access attempts by malware.  We note, the latest Android Jelly Bean 4.2 has an added security feature, Premium SMS Confirmation, that includes a giant list of premium shortcodes for each country and alerts a user anytime an app tries to send a message to a shortcode \cite{android42Security}. Our TWR permission model follows a similar approach but the security decision is based on presence or absence of gestures.

\subsection{Our Contributions}
%
The main contributions of this paper are summarized as follows.

\begin{enumerate}

\vspace{-1mm}
\item We propose TWR, a novel approach for malware prevention with an exclusive
focus on the smartphone platform based on intuitive gesture recognition. As
part of this system, we propose two novel light-weight gesture recognition
schemes that can be used in different contexts with little to no additional
user involvement.
The first mechanism, \textit{phone tapping detection based on accelerometer
data}, is geared for NFC applications, which usually require the user to tap
her phone with another device. The second mechanism, \textit{hand waving and
finger tapping / rubbing detection based on proximity sensor data}, involves
very simple gestures from the user, i.e., tapping or rubbing a finger near the
top of phone's screen or waving a hand close to the phone, and broadly appeal
to many applications (e.g., SMS). These gestures are as easy for users to
perform as many commonly deployed gestures, such as the iPhone's finger swiping
gesture.

\vspace{-1mm}
\item We outline how Tap-Wave-Rub can reside within the \textit{kernel-level
middle layer} between sensitive services and applications trying to access the
services, and be integrated specifically with the existing \textit{Android
permission model}.  This TWR-enhanced permission model provides continuous
enforcement of access control to sensitive resources and services even after an
application is installed on the platform. 

\vspace{-1mm}
\item We report on the implementation of our prototypes for the TWR gesture
recognition schemes on the Android platform. 

\vspace{-1mm}
\item To evaluate our approach, we conduct experiments to simulate the behavior of malware and normal user
usage activity. Our experiment results show that the proposed mechanism can
successfully detect malicious attempts to access sensitive services with high
detection rates, while imposing minimal usability burden.

\vspace{-1mm}
\end{enumerate}

The remainder of this paper is organized as follows. Section \ref{sec:related}
overviews related research. Section \ref{sec:ourapproach} presents the threat
model and design goals, and overviews our system. Section \ref{sec:design}
introduces the TWR enhanced permission model. Section \ref{sec:triggering}
presents two light-weight human gesture recognition schemes. Section
\ref{sec:experiment} shows the promising feasibility of our work through novel
experiments. Section \ref{sec:discuss} discusses various features of our system
and Section \ref{sec:conclusion} concludes the paper.


%% file: background.tex
\subsection{Malware Detection and Prevention}

Malware has threatened PCs for many years and a large number of malware defenses has accumulated in the literature. Generally speaking, the two most commonly used approaches for malware detection and analysis are
static analysis
\cite{christo2003static,venugopal2006efficient,shabtai2009detection} and
dynamic analysis
\cite{ellis2004behavioral,venugopal2007intelligent,bose2008behavioral,xie2010pbmds}.
Static analysis, also known as signature-based detection, is based on source or
binary code inspection to find suspicious patterns (malware) inside the code.
This approach has been used by many antivirus companies. However, it can be
evaded by malware authors through simple obfuscation, polymorphism and packing
techniques. Also it cannot detect zero day attacks. Dynamic analysis, also
known as behavior-based detection, monitors and compares the running behavior
of an application (e.g., system calls, file accesses, API calls) against
malicious and/or normal behavior profiles through the use of machine learning
techniques. It is more resilient to polymorphic worms and code obfuscation and
has the potential to defeat zero-day worms. 

Whether static or dynamic, current malware detection techniques for desktop
computers are still considered too time consuming for resource-constrained
mobile devices operated on battery. Most existing research focus on optimizing
desktop solutions to fit on mobile devices. The work of
\cite{venugopal2006efficient} tries to speed-up the signature lookup process in
static analysis by using hashes. Several collaborative analysis techniques have
been proposed to distribute the work of analysis by a network of devices
\cite{schmidt2009static,Shamili2010malware}. Remote server assisted analysis
techniques have also been proposed to reduce the overhead of computation on
individual devices \cite{cheng2007smartsiren,burguera2011crowdroid}.

Nevertheless, implementation of malware detection techniques on mobile phones
is still an emerging area of research and a challenging job
\cite{burguera2011crowdroid}. In practice, to protect mobile phones from
malware attacks, major mobile phone companies, such as Google, Apple, and
Nokia, use application review and permission model to prevent malware from being
installed at the first place. 

In the permission model, a user is asked to check permission requests of an application before installing it. The user either needs to grant all the permissions or choose not to proceed with the installation. So, it is the user's sole responsibility to decide whether the set of permissions granted to the application is potentially harmful or not. It requires users be well-educated to identify suspicious permission requests.  This practice of granting permissions all-at-once and only at installation time provides only coarse-grained access control to sensitive resources / services as subsequent permission check by the OS is transparent to the user. This leaves doors to malware writers as they can always expect users who are simply ignorant. Indeed, the widespread presence of mobile phone malware suggests that users actually do not care about giving permission to software they download. So the user is not aware how and when the permissions are exercised after the installation. Also permission model can be circumvented through permission escalation attacks \cite{davi2010privilege,felt2011permission}. Moreover, application developers do not always follow the least privilege rule with their permission requests. In a recent study, the authors pointed out one-third of application investigated are over privileged \cite{felt2011permission}.  So far, Android has the most extensive permission system and it is considered more vulnerable to potential malware attacks. 


The application review model adapts another approach.
Instead of relying on user permission, smartphone companies review applications
themselves before releasing them to people for download.  There are two types
of application reviews: manual and automatic. Apple reviews all applications in
the App Store while Symbian automatically tests Symbian signed applications. 
The manual process can be cumbersome and prone to human error while the
automatic review is not considered to be an effective deterrent
\cite{burguera2011crowdroid}.

The most closely related work to ours is the one proposed in
\cite{chaugule2011specification}. It shares similar philosophy as ours.  It
utilizes whether there are hardware interruptions to differentiate pure
software initiated action and human initiated action
\cite{chaugule2011specification}.  It aims at detecting malware specifically
targeting SMS and audio services.  These services usually start with user's
pressing or touching the keypad or touchscreen which generate hardware
interruptions for each key/screen press event. A purely software generated
activity (or malware generated activity), on the other hand, will not
explicitly generate a hardware interrupt. 
Although this approach is believed to be effective for malware detection, it
cannot help detect a more sophisticated malware such as the pickpocket
malware. This is because the pickpocket malware gets activated by user's
playing the tic-tac-toe game, which already involves touch screen activity that
can generate hardware interrupts.  The difference between
\cite{chaugule2011specification} and our work can be summarized as
follows.  \cite{chaugule2011specification} attempts to check whether there is
(any) user activity whereas our goal is to check whether there is a \textbf{special user-aware} activity. So our approach provides more fine-grained access control to sensitive services and thus can detect even sophisticated malware.


Another work that parallels to ours was recently presented in
\cite{franzi2012udac}. It proposes an approach of user-driven access control by
granting permission to the application when user's permission granting intent
is captured. It introduces access control gadgets (ACGs) which are UI elements
exposed by each user-owned resource for applications to embed. The user's
authentic UI interaction with corresponding ACG grant the permission to an
application to access the corresponding resource. A fundamental difference
between \cite{franzi2012udac} and our work is that the design proposed by
\cite{franzi2012udac} grants the permission to an application when user's
authentic \textbf{UI interaction} with corresponding ACG is captured whereas
our design grants permission to an application when a specific user
\textbf{gesture} (tapping/waving) is captured. The design proposed by
\cite{franzi2012udac} not only requires \textit{kernel level changes} but also
necessitates \textit{application level modifications}. It also requires
Resource Monitor (RM) to be incorporated for each resource such as the device
drivers. Moreover, it requires additional composition ACG (C-ACG) along with
composition RM if an application requires different resources to be
accessed/used. Our work, in contrast to \cite{franzi2012udac}, has an advantage
in that it neither requires application level changes nor requires resource
monitor to be added for each resource. Note that if there are many resources
that can be used by an application, then the number of C-ACG and C-RM will
become extremely large. Another advantage of our work over
\cite{franzi2012udac} is that our design supports ``services'' (such as NFC)
which do not have any specific UI elements or ACGs associated with them.  For
the approach of \cite{franzi2012udac} to work with services like NFC,
additional ACG for UI interaction will need to be added, which will
significantly hamper the usability of such services. In contrast, in our case,
implicit permission granting intent is acquired by capturing the tapping
gesture.

\subsection{Human Activity Detection and Security}
Gesture recognition has been extensively studied to support spontaneous
interactions with consumer electronics and mobile devices in the context of
pervasive computing \cite{baudel1993charade,cao2003visionwand,uwave}. Due to
the uniqueness of gestures to different users,  personalized gestures have been
used for various security purposes. 

Gesture recognition has been used for user authentication to address the
problem of illegal use of stolen devices
\cite{gafurov2006biometric,conti2011mind}.  In \cite{gafurov2006biometric},  a
mobile device gets unlocked to use when it detects the gait (walking pattern)
of the legitimate owner. In \cite{conti2011mind}, a smartphone
gets locked when it does not detect the ``picking-up phone'' gesture which the
owner naturally performs to answer a phone. Both works provide transparent user
authentication and do not require explicit user involvement.
\cite{liu2009user} reports a series of user studies that evaluate the
feasibility and usability of light-weight user authentication based on gesture
recognition using a single tri-axis accelerometer.

Gesture recognition is also used to defend against unauthorized reading and
Ghost-and-Leech relay attacks in RFID systems
\cite{secret_handshakes,halevi2012sensing}. The secret handshake scheme
proposed in \cite{secret_handshakes} allows an RFID tag to respond to reader
query selectively when the tag owner moves the tag in a certain pattern (i.e.,
secret handshake). The work of \cite{halevi2012sensing} uses posture as a valid
context to unlock a tag in some RFID applications without changing the
underlying user usage model.

The use of unique key press gestures or secure attention sequences (SASs), such
as CTRL-ALT-DEL, may also serve as a means to defend against malware. However,
we are not aware of SASs being currently used on mobile phones. SASs need to be
unique and usually require multiple key presses simultaneously (e.g.,
CTRL-ALT-DEL). Such sequences will be very hard for the user to perform on
phones. Some of the gestures proposed in our paper (i.e., hand waving, finger
tapping or rubbing) can be viewed as a form of novel and user-friendly SAS
suitable for phones. 
To distinguish a malware activy from a human user activity,
\cite{gummadi-notABot} rely upon detecting keyboard or mouse activity and a
Trusted Platform Module (TPM). However, this approach is limited to devices
equipped with a TPM, and thus is not suitable for most (if not all) current
mobile phones and smartphones. 

%% file: threat.tex
\subsection{Threat Model} \label{sec:threat}
%



In our model, we assume that the mobile phone is already infected with malware.
As in the pickpocketing attack of \cite{epickpocket}, the malware could reside
within a benign looking application (e.g., a game) which the user may have
downloaded from an untrusted source. Our model covers a broad range of malware
and does not impose any restriction on malware behavior except that an action
from the malware is not human-triggered.  For example, the malware may want to
access a service or resource (such as NFC, SMS or GPS) available on the phone
itself, or to communicate with an external entity, such as an
attacker-controlled remote server (botmaster).

We assume that the OS kernel is healthy and immune to malware infection. In particular, the malware is not able to maliciously alter the kernel control flow. Also, the phone hardware is assumed to be malware-free.
Specifically, we assume that the malware can not manipulate the input to, and output from, the phone's on-board sensors. 

We do not impose any restriction as to how frequently the malware attempts to
access a given service. However, in order to remain stealthy, constantly
attempting access would not be feasible for the malware, and rather random or
periodic sampling is expected. 

In addition to the user space level control of the phone, the malware may
collude and synchronize with an entity in close physical proximity of the phone
(and its user).  This external entity may attempt to manipulate the physical
environment in which the phone is present or interfere with the user per se. 
We do not, however, allow this attacker to have physical access to the phone.
That is, if the attacker has physical access to the phone, then he can
lock/unlock a resource just like the phone's user. In other words, our
mechanisms are not meant for user authentication and do not provide protection
in the face of loss or theft of phone.  

%% file: design.tex
\subsection{Design Goals} \label{sec:designgoals}
For our malware prevention approach to be useful in practice, it must satisfy
the following properties:

\begin{itemize}
\vspace{-1mm}
\item \emph{\textbf{Lightweight-ness:}} The approach should be lightweight
in terms of the various required resources available on the phone, such as memory, computation and battery power.  

\vspace{-1mm}
\item \emph{\textbf{Efficiency:}} The approach should incur little
delay. Otherwise, it can affect the overall usability of the system. We believe
that no more than a few seconds should be spent executing the approach. 

\vspace{-1mm}
\item \emph{\textbf{Robustness:}} The approach should be tolerant to
errors. Both the False Negative Rate (FNR) and False Positive Rate (FPR) should
be quite low. A low FNR means that a user would, with high probability, be able
to execute an application (which accesses some sensitive services) without
being rejected. Low FNR also implies a better usability.  On the other hand,
Low FPR means that there should be little probability to grant access to a
sensitive service when a user does not intend to do so. Low FPR clearly implies
a little chance for malware to evade detection.

\vspace{-1mm}
\item \emph{\textbf{Usage Model Consistency:}} The solution should
require little, or no change, to the usage model of existing smartphone
applications. Ideally, if the use of a particular phone service can be commonly
associated with a particular (unique) gesture (e.g., phone tapping for NFC),
this gesture may be used to specially protect the said service. In this case,
no changes to the adopted usage model will be necessary.  It is also possible
that there is no unique gesture pattern that can be found to use a certain
service (e.g., Bluetooth). In such a situation, an intuitive gesture template
can be associated with that service and a user will be required to explicitly
perform the hand movements defined by that gesture. In this case, only minor
changes to the adopted usage model will be imposed.
\vspace{-1mm}
\end{itemize}

%% file: overview.tex
\subsection{Why Tap-Wave-Rub?} \label{sec:overview}
At a higher level, to distinguish between malware and human-initiated activity,
we envision two different categories of methods: \textit{implicit} and
\textit{explicit}. The implicit approaches automatically infer whether or not a
human activity is taking place based on contextual information.  The explicit
approach, on the other hand, determines the presence of human user by means of
explicit human involvement.  








%


A number of context-aware malware detection mechanisms are possible. The
implicit category of mechanisms could be applied in scenarios whereby the
access to a resource is pre-empted by a specific human gesture. It would
certainly be quick and user-friendly since no additional work is needed by the
user. The use of keypad or touchscreen activity, as previously studied in
\cite{chaugule2011specification}, might be a good fit in some scenarios, such
as when accessing SMS and audio services. These services usually start with
user's pressing or touching the keypad or touchscreen which generate hardware
interruptions for each key/screen press event. A purely software generated
activity (or malware generated activity), on the other hand, will not
explicitly generate a hardware interrupt. 
Although this approach might be effective to prevent malwares targeting SMS and
audio services, it can not help prevent the activation of a more sophisticated
malware such as the NFC pickpocket malware \cite{epickpocket}. It is because this
malware gets activated by user's playing the tic-tac-toe game, which already
involves touch screen activity that can generate hardware interrupts. In other
words, the use of keypad or touchscreen activity, as a context for malware detection, is expected to result in high false unlocking likelihood, and thus poor security. 

Fortunately, in the context of NFC, there already exists a natural gesture,
``phone tapping'' using which NFC services are accessed.
Tapping involves touching the phone against an RFID tag, or another NFC device,
and is a gesture users commonly need to perform to 
use the NFC functionality. In this paper, we develop a \textit{phone tapping
detection} mechanism based on accelerometer data. As we would
demonstrate in the subsequent sections, this mechanism satisfies all of our
design goals, being lightweight, efficient, robust and user-friendly.

Unlike NFC, most other services/resources on the phone may not be associated
with a unique implicit gesture. In such situations, an explicit human
involvement would be necessary. At first glance, Biometrics as well as
human-interactive proofs, e.g., CAPTCHAs, appear to be natural candidates for
such an explicit detection. Both will prove that a human is the one initiating
an access request. Biometrics, in addition, also prove that this human is the
owner of the phone, and thus provide protection in the event of loss/theft of
device (this level of protection is beyond our threat model though). A variety
of biometrics (such as face or voice recognition) are already available on
commodity phones. Similarly, CAPTCHAs (such as textual/aural/visual ones) are
ubiquitous on the Internet and can be easily adopted on the phones.
However, both of these mechanisms do not satisfy many of our design goals.
Biometrics are not lightweight, can be time consuming, have been known to be
error prone, and not very user-friendly (and rather invasive). CAPTCHAs, on the
other hand, are reasonably lightweight (especially those based on text), but
long known to be a source of user frustration on the Internet, perhaps more so
on the phones due to their small form factors complicating both display of the
challenge and entry of the response.   

Since CAPTCHAs and biometrics do not seem to be satisfactory means of explicit
malware detection, we turn our attention to lightweight and natural gestures.
In particular, we explore hand waving, finger tapping or rubbing, all of which
could be identified by proximity sensors available on most modern phones.  A
wave involves waving a hand in close proximity of the phone. A tap/rub involves
tapping or rubbing a finger near the top of phone's screen, which is where the
proximity sensor is usually located. All these gestures cause a sudden
fluctuation in the readings of the proximity sensor, and can be very easily
identified, and are robust to errors, as we will demonstrate in the following
sections.  It is important to note that the proximity sensor is located off of
the touch screen or display, and therefore regular touchscreen inputs are not
likely to trigger a tap or rub. Wave is also a unique gesture quite unlikely to
be exhibited in daily activities of users.  





%% file: systemmodel.tex

\section{TWR-Enhanced Permission Model} \label{sec:design}

Permission models have become very common on smartphone operating systems to
provide access control to sensitive services for installed third party
application. The Android platform has the most extensive permission system and
poses to become a market leader. Thus, we base the design of our TWR system on
the Android platform.

The idea of the TWR system is to add another layer of permission check before the original
Android permission check. 
As stated in Section \ref{sec:threat}, we assume the adversary is not able to maliciously alter the kernel control flow. So gesture detection forms a trusted path with the OS.

\begin{figure}[h]
\centering
\includegraphics[width=1.1\columnwidth]{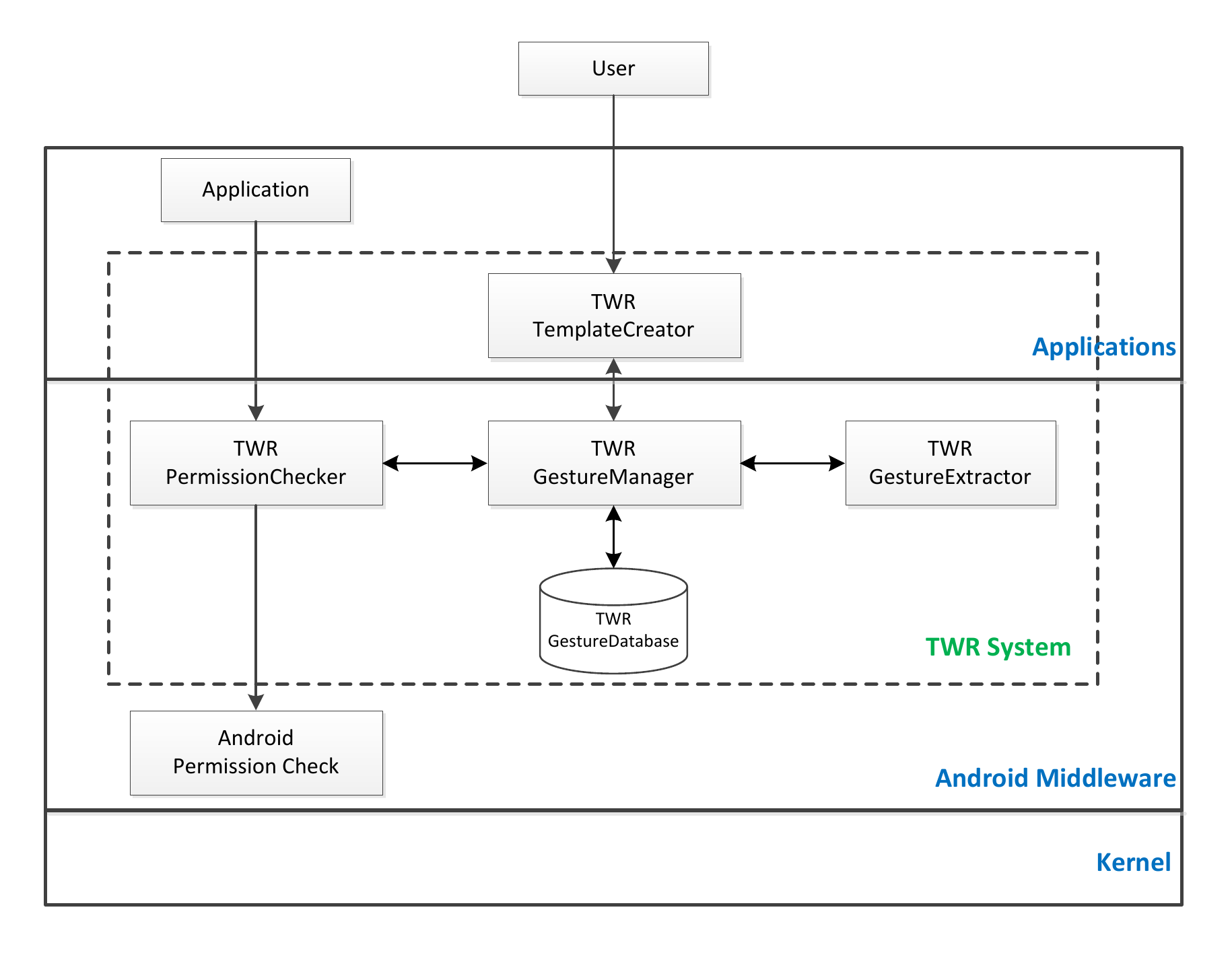}
\caption{The TWR Architecture}
\label{fig:architecture}
\end{figure}

Intercepted permission requests are handled by the
five components in the TWR's architecture: \emph{TWR PermissionChecker},
\emph{TWR GestureManager}, \emph{TWR GestureExtractor}, \emph{TWR
TemplateCreator}, and \emph{TWR GestureDatabase}. The architecture of TWR is
depicted in Figure \ref{fig:architecture}. In the following paragraphs, we present the role of each TWR component by describing the possible interaction between them and the outside world.

The \emph{TWR PermissionChecker} stands in front of the original Android Permission
check. When an application initiates a request to access a sensitive service,
the request is intercepted by \emph{TWR PermissionChecker}. This component
interacts with \emph{TWR GestureManager} to check whether the requested service
is protected by a certain gesture. If not, the request is forwarded to the
Android Permission Check as usual. Otherwise, \emph{TWR GestureManager} interacts with the \emph{TWR GestureExtractor} to begin collecting gesture data (tapping, rubbing, or waving in this paper). The captured data is then sent to the \emph{TWR GestureManager} for further process.

Here we distinguish between two types of gesture recognition: user-dependent and
user-independent. As their names suggest, a gesture is user-dependent if there is significant variation among gesture data from multiple participants for the same predefined gesture; while a gesture is user-independent if either there is no apparent difference or the recognition process does not differentiate among
user data from multiple participants. Our phone tap gesture recognition is
user-dependent as it captures the user's own features of the tapping movement.
Given that users can hold a phone in different ways and use different forces to tap, the phone tap gesture is thus user-dependent. Our hand wave or finger tap/rubbing recognition scheme is
user-independent as it infers user activity by checking whether a special
location (in our context, the place where the proximity sensor is located) is
touched or not (instead of the potentially biometric feature of human movement). 

For user-dependent gesture recognition, we usually need to create a gesture
template which is used as a reference in the actual recognition stage. The user
can interact with the \emph{TWR TemplateCreator} to register a new gesture
template, to update and delete exiting gesture templates. \emph{TWR
TemplateCreator} is an Android application which allows interaction between TWR
and the user. When the user creates, deletes, or modifies the gesture
information, it needs to retrieve and store the information to \emph{TWR
GestureDatabase} via \emph{TWR GestureManager}. \emph{TWR GestureManager} is
the only component that has access to \emph{TWR GestureDatabase}. 

So depending on the type of gesture recognition scheme (user-dependent or
user-independent), the \emph{TWR GestureManager} processes the gesture data
from \emph{TWR GestureExtractor} differently. If the gesture is user-dependent,
it compares the similarity between the newly captured data with the
corresponding gesture template stored in the \emph{TWR GestureDatabase}. If the
gesture is user-independent, the \emph{TWR GestureManager} determines directly
whether a gesture is performed or not by utilizing information in the captured
data without the help of a template. In either case, if a required gesture is
detected, the request is forwarded to Android Permission Check for further
check. Otherwise, the request for service access is rejected.

%% file: tapping.tex
\subsection{Phone Tap Detection} \label{sec:tapping}
As we mentioned in Section \ref{sec:related}, the use of hardware interruption
to differentiate between pure software initiated activity and human initiated
activity is not effective to prevent malware hidden under the cover of a
victim app, since activating the victim app already involves keypad click or
screen touch which can generate hardware interrupts. This motivates us to use
app-specific user events to distinguish between hidden malware and an app
initiated by a human being. That is, instead of simply using general key/screen
press events to infer human activity, we try to recognize whether it involves
the \emph{right} activity a user needs to do to access a sensitive service,
such as the access to NFC. 

A smartphone is a personal hand-held device installed with a lot of apps. Most
of the time, these apps are activated by specific phone/hand movements. For
example, when a user wants to place a call, she needs to unlock the screen,
activates the phone app, inputs the number (or clicks on a name in the contact
list), and then puts the phone near the ear to start the call. Also, to use the
NFC to scan a smart poster, a user needs to unlock the screen, activates the
NFC reader app, and taps the phone on the smart poster to read information.
Since ``tapping'' (touching the phone against an RFID tag, or another NFC
device) is a gesture which users commonly need to perform to use the NFC
functionality, as an illustrative example, we can use tapping to determine
whether an NFC access is human-initiated or not. Intuitively, tapping on a
smart-poster should be different from other user phone activities (such as
keyboard click or screen touch) and user physical activities (such as walking
or running). 

One advantage of this tapping approach is that it does not require any
additional user activity besides what is being used commonly, and thus transparently recognizes user activity when a user taps a smart poster to obtain information. 
However, it may exhibit false positive rates and not fully prevent the
pickpocket malware activation since normal user activities (such as playing the
game) may generate motions similar to tapping. To achieve higher prevention
rate, we can try other intuitive user-aware gestures similar to tapping, such
as ``tapping twice'' or ``tapping thrice'' in succession.

To recognize tapping, we utilize the on-board accelerometer data. An
accelerometer sensor measures the forces applied to the phone (minus the force
of gravity) on the three axes: $x$, $y$, and $z$. Let $(a_x, a_y, a_z)$ denote
the values corresponding to the 3 axes from the accelerometer. 

Our detection algorithm consists of two phases: training phase and recognition phase. In the
training phase, a user performs the target action (tapping) multiple times, and
accelerometer data of the action is recorded and processed to generate a
tapping template. The template serves as a reference to be compared later with
real-time user movement data: a match indicates the recognition of user tapping
activity; otherwise, it is inferred that either there is no user activity or
the activity is not the ``valid'' user activity to grant NFC access.

After the training phase, the system compares a newly observed movement with
the template. The system records the accelerometer data, from the moment the
user activates the NFC reader app, until she taps on a smart poster. To
recognize tapping, the system computes the cross-correlation $C$ of the
acceleration data $A$ against the template $T$, both of size $n$ data points as shown in Equation \ref{eq:cross-correlation}.

The cross-correlation $C$ computed from Equation \ref{eq:cross-correlation}
when comparing two time series is a real value, representing a similarity
measure. The higher the value of $C$, the higher the similarity between the two series.
The maximum value is obtained when the two series under comparison are
identical. A movement is considered a valid tapping activity when the computed
cross-correlation $C$ exceeds a certain cross-correlation threshold which is
usually obtained through empirical study. 

\begin{align} \label{eq:cross-correlation}
C(A,T) &= \frac{\sum_{i=1}^n(a_{x_i}-{{\bar{a}_x}})(T_{x_i}-{\bar{T}_x})}{\sqrt{\sum_{i=1}^n(a_{x_i}-{\bar{a}_x})^2}\sqrt{\sum_{i=1}^n(T_{x_i}-{\bar{T}_x})^2}} \nonumber\\ 
&=\frac{\sum_{i=1}^n(a_{y_i}-{\bar{a}_y})(T_{y_i}-{\bar{T}_y})}{\sqrt{\sum_{i=1}^n(a_{y_i}-{\bar{a}_y})^2}\sqrt{\sum_{i=1}^n(T_{y_i}-{\bar{T}_y})^2}} \nonumber\\
&=\frac{\sum_{i=1}^n(a_{z_i}-{\bar{a}_z})(T_{z_i}-{\bar{T}_z})}{\sqrt{\sum_{i=1}^n(a_{z_i}-{\bar{a}_z})^2}\sqrt{\sum_{i=1}^n(T_{z_i}-{\bar{T}_z})^2}}
\end{align}
where $\bar{a}_x$ denotes the means of time series $a_{x_i}$ for $i \in [1,n]$ (others follow the same notation).

Here we describe one way to determine the cross-correlation threshold. Suppose we have $m$ traces of tapping movements $T_1$, ..., $T_m$. The threshold $C_T$ can be estimated as the minimum cross-correlation between any two series $T_i$ and $T_j$ ($i, j \in [1, m]$ and $i \neq j$). That is:
\begin{equation} \label{eq:threshold}
C_T = min_{i,j=1}^m(C(T_i,T_j))
\end{equation}




%% file: waving.tex
\subsection{Hand Wave, Finger Tap or Rub Detection} \label{sec:waving}

\begin{figure*}[ht]
\begin{center}
\begin{minipage}{0.3\linewidth}
\includegraphics[width=3.2cm]{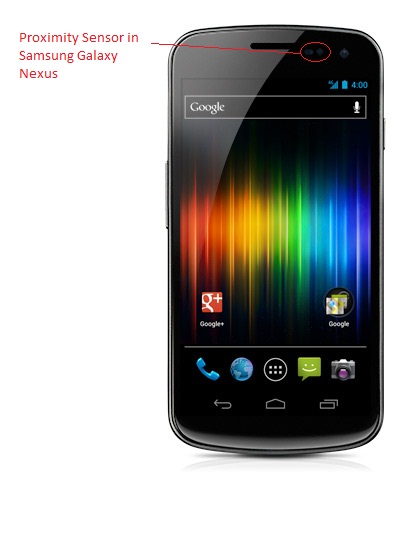}
\vspace{-3mm}
\label{p1}
\end{minipage}
\begin{minipage}{0.3\linewidth}
\includegraphics[width=4.2cm]{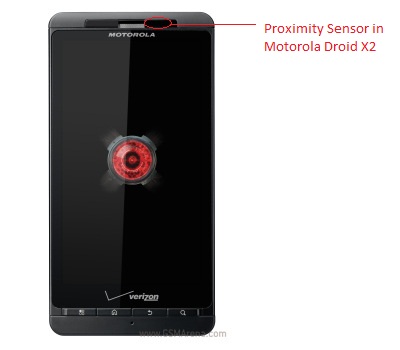}
\label{p2}
\end{minipage}
\begin{minipage}{0.3\linewidth}
\centering
\includegraphics[width=4.1cm]{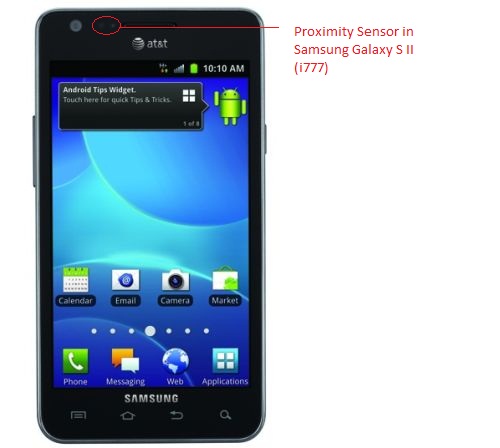}
\label{p3}
\end{minipage}
\vspace{-3mm}
\caption{The locations of proximity sensor on a sample of smartphones (Galaxy Nexus, Droid X2, Galaxy S II)}
\label{fig:loc-prox}
\end{center}
\end{figure*}

The goal of a proximity sensor is to detect the presence of a nearby object, not
necessarily in physical contact with the sensor. This sensor works by sending
an electromagnetic signal and analyzing the change in the electromagnetic field
or the returned signal itself. Usually, the range of a proximity sensor is very
small (few cms at most), and different forms of proximity sensors have been
invented \cite{prox-sensor}. Such sensors have seen an ubiquitous deployment on
current mobile phones. For mobile phone manufacturers, the primary use of a
proximity sensor is to help preserve battery power. This is achieved by turning
off the phone's display especially in cases when the user is on a phone call
where user's ear is close to the proximity sensor. This is the reason the
proximity sensor is located near the ear piece of a mobile phone, off of its
display (see Figure \ref{fig:loc-prox}). This is true for most, if not all,
smartphones, including the Androids and iPhones.  We note that this is a
property that we carefully leverage in developing our tapping or rubbing
gesture mechanisms for malware defense. Such gestures do not interfere with the
gestures made by the user while interacting with the phones' (touch screen)
display, and significantly reduce the False Positive Rate (FPR).

In order to utilize the proximity sensor for our purpose, we needed a human
gesture that can ``trigger'' the sensor in some way and not likely to be
exhibited in daily activities. To this end, we resort to gestures that can
fluctuate the readings of the sensor quickly for a short duration of time. This
could be easily achieved by the user bringing in, and moving out, her
hand/finger in close proximity of the sensor. This inspired us to develop three gestures:
hand waving (waving one's hand close to the phone), finger tapping (tapping
a finger near the sensor, i.e., close to the top of the screen) and rubbing
(rubbing a finger near the sensor). 

The algorithm to detect quick fluctuations in the reading of the proximity
sensor is very simple and straightforward (which makes it lightweight, satisfying one of
our design goals). Basically, when there is a proximity change, the time is
recorded and the difference in time reading with an $n$-th previous reading (we
use $n=6$) is compared. If the difference is less than a pre-specified time
limit (the upper bound of the duration of the gesture; in our case 1.5 seconds),
the device will be unlocked. Otherwise, it will remain locked. A detailed
pseudocode for this simple procedure is outlined in Algorithm \ref{pseudocode}.

\begin{algorithm}[h]
\caption{\bf Tap-Wave-Rub Detection using Proximity Sensor Data}
\begin{algorithmic}[1]
{\small{
	\STATE Set $proxIndex=0$, $WIND\_SZ=6$, $WAVE\_TIME\_LIMIT=1500~ms$, and $UNLOCK\_TIME\_FRAME=1000~ms$
	\STATE Record the time whenever proximity sensor detects a change
		\begin{enumerate}
		\item An array $ProxChangeTime$ with size equal to $WIND\_SZ$  was used to record the time of proximity sensor change in cyclic order.
		\end{enumerate}
	\STATE Calculate the time difference between the current time with the time recorded in previous six sensor change value.
		\begin{enumerate}
		\item $TimeDiff = ProxChangeTime[proxIndex] - ProxChangeTime[(proxIndex+1) \% WIND\_SZ]$
		\end{enumerate}

	\STATE If TimeDiff is less than $WAVE\_TIME\_LIMIT$, Unlock for $UNLOCK\_TIME\_FRAME$
	\STATE Increase $ProxIndex$ by 1, i.e., $proxIndex = (proxIndex+1) \% WIND\_SZ$
	\STATE Repeat Step 2.
}}
\end{algorithmic}
\label{pseudocode}
\end{algorithm}

%% file: setup.tex
\subsection{Prototypes and Test Devices} \label{sec:setup}

To evaluate the feasibility of the TWR gesture detection mechanisms, we first
developed prototype applications on the Android platform. The phone tapping
detection scheme was implemented and installed on a Google Nexus S Android
phone, while the hand waving, finger tapping/rubbing detection scheme was
implemented and installed on both a Droid X2 phone and a Samsung Galaxy Nexus
(I9250) phone. The two Nexus phones come equipped with NFC chips, and are
therefore a good target devices for an eventual deployment of our approach. 

%% file: tap-experiment.tex
\subsection{Phone Tap Experiment} \label{sec:tap-exp}

In this section, we report on evaluation of tapping based user activity
recognition scheme outlined in Section \ref{sec:tapping}. This scheme is
specially designed to protect against malware targeting NFC reading services, 
since tapping is a natural hand movement which a user needs to perform to use
the reader function of NFC. 


Since ``tapping'' (touching the phone against an RFID tag, or another NFC
device) is a very simple hand movement, we hypothesize it might be confused
with other user movements such as those users perform when they play games,
and thus have higher false positive rate, FPR (or lower prevention rate). In a hope to
achieve higher prevention rate, we also experiment with two other intuitive
user-aware gestures similar to tapping: tapping twice and tapping thrice in
succession. We call these three tapping gestures as ``tapping once'', ``tapping
twice'', and ``tapping thrice'', respectively.

First, to determine the cross-correlation detection threshold, we collected 30
traces of accelerometer data for each tapping gesture.  Each of our trace
contains 100 data points and is recorded over a 2-second time period (we wanted
our schemes to be efficient). Each trace is then used as a template, which is
compared with all the other 29 traces to calculate a serial of $C$ values. The
smallest $C$ value is chosen as the threshold value.  This threshold value is
then stored with the corresponding tapping template and a matched posture needs
to yield a $C$ value larger than this threshold.  These traces were collected
by the experimenter while performing NFC tapping gesture 30 times. Such data
collection and testing methodology is in-line with related prior security work,
e.g., Secret Handshakes \cite{secret_handshakes}. 
Our methodology captures a realistic usage scenario whereby each user can be
trained ``once by their phone'' and can create their template, e.g., when they
purchase the phone.

%
\begin{table*}[t]
{\small{
\vspace{-3mm}
\caption{Phone Tapping Detection Results (rates at which a gesture shown on each row matches with gesture/activity shown on each column}
\begin{center}
\begin{tabular}{|c|c|c|c|c|c|c|c|c|} 
\hline
 & \textbf{Tapping} & \textbf{Tapping} & \textbf{Tapping} & \textbf{Walking} & \textbf{Walking} & \textbf{Still} & \textbf{Screen-touch}& \textbf{Phone}\\ 
 & \textbf{Once} & \textbf{Twice} & \textbf{Thrice} & & \textbf{Stairs} &  & \textbf{Activities}& \textbf{Movement}\\ 
[0.5ex] 
\hline\hline 
\textbf{Tapping Once} & 94.67$\%$  &  NA  &  NA  &  0$\%$ &   0$\%$ & 0$\%$ &   0$\%$ & 2$\%$ \\\hline
\textbf{Tapping Twice} & NA  &  92.67$\%$  &  NA  &  0$\%$ & 1.33$\%$ &   0$\%$ & 0$\%$ &   0$\%$ \\\hline
\textbf{Tapping Thrice} & NA  &  NA & 96.67$\%$  &  0$\%$ & 5.33$\%$ &   0$\%$ & 0$\%$ &   0$\%$ \\\hline
\end{tabular}
\end{center}
\label{tab:tap-exp} 
\vspace{-2mm}
}}
\end{table*} 

We first test the performance of the three tapping gestures to identify which
one can have higher recognition rate (thus lower false negative rate, FNR).  To do
this, we collect a total of 150 traces for each tapping gesture, 30 traces every
day for 5 days. We then use the template and the threshold calculated above to
determine the recognition rate. The successful recognition rate is listed, in the form of a confusion matrix, in
Table \ref{tab:tap-exp}. It shows that ``tapping once'' achieves high
recognition rate $94.67\%$ (or a low FNR $5.33\%$) compared to the other two tapping gestures. 

We next test the performance of the three tapping gestures to identify which
one is the least to be confused with other user or phone movements and thus has
low FPR. It is important to evaluate the FPR. If a tapping gesture can be very
similar to a certain other movement (accidental or manipulated by an attacker),
the malware may circumvent the gesture detection process. 

To determine the FPR, we compare tapping postures with many phone/user movements. These movements might be just normal user activities, or activities coerced by a nearby attacker. They include user movements such as:
walking, walking stairs, screen-touch activities (text messaging and surfing
Internet), phone-moving activity (motion gaming and picking up calls), as well
as, the scenarios where phone is left still. For each movement, we also collect a total of
150 traces, 30 traces every day for 5 days. The error rates are all 
listed in Table \ref{tab:tap-exp}. 

Our experiment result shows ``tapping once'' is very unlikely to be confused
with walking, walking stairs, still, and screen-touch activities such as text
messaging or Internet browsing. However, it might occasionally be confused with
phone motion caused when the user plays game or picks up a phone call with a
false positive rate of $2\%$. ``tapping twice'' and ``tapping thrice'', on the
other hand, are very resilient to phone motions but they resemble motions when
a user walks on stairs. Nevertheless, all achieve satisfying low false positive
rate. 

One potential reason why the false positive rate is low  might be that tapping is a type of user-aware movement. When performing such a gesture, the user is
believed to be aware of her hand movement. So gestures are performed in a
more-or-less controlled way, e.g., the phone is always held in the similar way
when a user performs tapping. In non-user-aware movements, on the other hand,
the phone can be tilted in any position. The reference template is usually
collected in a reference coordinate system. However, once the phone is tilted,
movement data collected from the device is no longer in the reference
coordinate system and the corresponding movement will not be detected
correctly. In this way, user-aware gesture is very unlikely to be similar with
user-unaware movements, and thus has low false positive rate. Previous studies
on gesture recognition also suggest certain gestures can be quite unique
and different from other gestures \cite{secret_handshakes,uwave}. So tapping can be distinguished from other user-aware movements such as ``picking up the call''. 

Our experiment result, contrary to our hypothesis that ``tapping once'' may
have high false positive rate, shows that ``tapping once'' actually achieves
both high recognition rate and low false positive rate, and has a performance
comparable to ``tapping twice'' and ``tapping thrice''. However, ``tapping
once'' outperforms the other two tapping gestures in term of efficiency, and
has superior usability since it does not require any change in the user usage
model of NFC.

%% file: wave-experiment.tex
\subsection{Hand Wave, Finger Tap and Rub Experiment} \label{sec:wave-exp}

\begin{table*}[t]
{\small{
\vspace{-3mm}
\caption{Hand Waving, and Finger Tapping / Rubbing Detection Results (rates at which a gesture shown on each row matches with gesture/activity shown on each column)}
\begin{center}   
\begin{tabular}{|c|c|c|c|c|c|c|c|c|c|} 
\hline
 & \textbf{Hand} & \textbf{Tapping /} &  \textbf{Walking} & \textbf{Phone Drop/} & \textbf{Daily} & \textbf{Screen-touch}& \textbf{Game}&\textbf{Game} & \textbf{Bumping}\\ 
 & \textbf{Waving} & \textbf{Rubbing} &  & \textbf{} \textbf{Fall}&  \textbf{Activity}& \textbf{Activities}& \textbf{Play (O1)}&\textbf{Play (O2)}&\\ 
[0.5ex] 
\hline\hline 
\textbf{Hand Waving } & 93.75$\%$  &  NA  &  0$\%$ &   0$\%$ & 0$\%$ &   0$\%$ & 0$\%$  &0$\%$ &0$\%$ \\\hline
\textbf{Tapping / Rubbing} & NA  &  96.25$\%$  &  0$\%$ & 0$\%$ &   0$\%$ & 0$\%$ &   0$\%$& 6.5$\%$ &0$\%$ \\\hline
\end{tabular}
\end{center}
\label{tab:prox-exp} 
\vspace{-2mm}
}}
\end{table*}

We conducted several experiments to evaluate our prototype implementing the
wave-tap-rub detection mechanism for malware detection. As in the case of the
phone tapping experiments,  the goal of our tests was to primarily estimate the
error rates, i.e., FNR and FPR. 

In order to determine FNR, one of the authors of this
submission attempted to unlock the phone using the hand waving as well as
rubbing gesture. 50 trials using each gesture were performed spanning over a
period of 2 days. Out of these, only on 1 occasion, this user failed to unlock
the phone, resulting in an average recognition rate of 98\% (or FNR of 2\%). Since multiple
trials may have trained this user significantly likely leading to a bias, we
further conducted our tests with multiple other users. These volunteers were
drawn from our Department (CS) and were mostly students at undergraduate and
graduate levels. The users were first explained the purpose of the study and
then demonstrated the gestures using which they were to unlock the phone. The
users were specified the location of the proximity sensor on the phone.
Although in real life, users may not be aware of the location of the proximity
sensor, they can be easily provided with this information using a simple
interface. For example, an arrow pointer could be provided on the screen which
points to the proximity sensor, and user could be asked to wave/tap/rub
accordingly.  

A total of 16
volunteers participated in our study. 
8 of the participants were requested to perform
the waving based unlocking 10 times and the results were recorded automatically by our
program. 
The resulting average recognition rate observed was 93.75\% (or FNR of 6.25\% [5/80]).

The other 8 users were asked to evaluate the rub gesture or finger tapping
based unlocking.  The users were not restricted to using one or the other
gesture, and it was up to their discretion as to which one they wanted to use.
The observed average recognition rate in this case was 96.25\% (or FNR of 3.75\% [3/80]).

These error rates can be deemed to be fairly low, and we expect them to further
reduce significantly as users become more and more familiar with such gestures.



Next, we set out to evaluate the likelihood of false unlocking
under different activities. These activities might be just daily user activities, or activities coerced by a nearby attacker.
To this end, the experimenter conducted several tests emulating different user
activities that have the potential of triggering the proximity sensor
fluctuations. In each test, the phone was set to record a trigger event and its
frequency. Each of the test was performed with each of our test devices (Droid and Nexus).

First, to simulate a walking activity, the mobile phone was stowed in a
backpack and the experimenter, carrying the backpack on shoulders, walked around
for 10 minutes. No trigger events occurred in this case. The experiment was
repeated at a later point of time for a duration of 15 minutes.
Still, the phone did not get unlock at all. We further continued this
experiment for a duration of 2 hours during which the experimenter was walking
and driving a car. No unlocking events occurred even in this case also.

To trigger sudden fluctuations to the proximity sensor readings, we next
conducted a ``drop and fall'' test. This mimicked a situation where the phone
accidentally drops on, or is thrown at, a surface. Clearly, we could not just
drop or throw our test device on the floor to avoid damaging it.  To do this
meaningfully, therefore, we first threw our test device on a bed, with the
screen (and thus the proximity sensor) facing the bed, from a height of 50cm to
70cm. 10 trials of this test were performed. Then, we similarly dropped the
device on a couch 15 times and a bed 15 times. No triggering events were
recorded over this set of tests.  

Our next test involved treating the test device as the experimenter's own
device. During this test, which last for about 1 day, the experimenter made and
received phone calls, typed SMS messages, browsed the Internet, walked around
while the phone was kept in pocket, went up/downstairs, drove, and left the phone on a table near a
keyboard while typing on the keyboard, among other chore activities. No
unlocking was registered. 

In the above test, we had evaluated the effect of touchscreen typing on the
phone while writing SMSs and browsing the Internet. Since the proximity sensor
is located on top of the screen, such touchscreen activities were naturally not
supposed to be affecting the sensor's output. However, more involved
touchscreen activities, such as game play, may have an impact and may mimic the
rub gesture. 

\begin{figure}[h]
\begin{minipage}[b]{\linewidth}
\centering
\includegraphics[width=8.5cm]{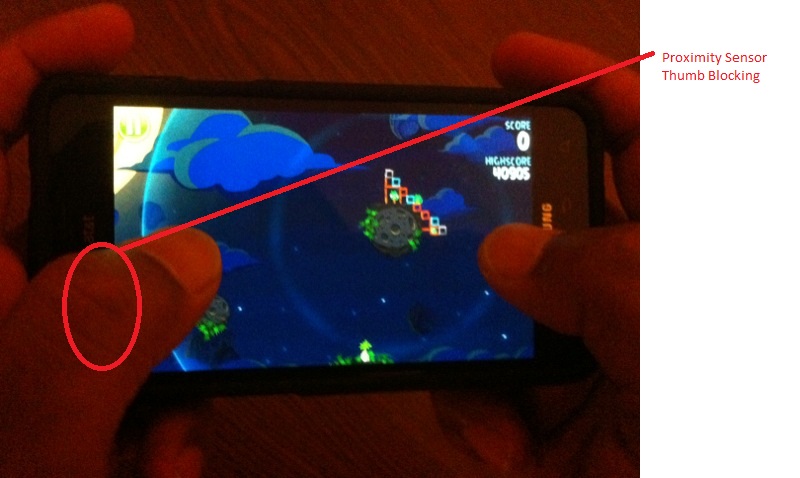}
\caption{Orientation 1 (O1): Thumb Blocking the Proximity Sensor}
\label{o1}
\end{minipage}
\begin{minipage}[b]{\linewidth}
\centering
\vspace{4mm}
\includegraphics[width=8.5cm]{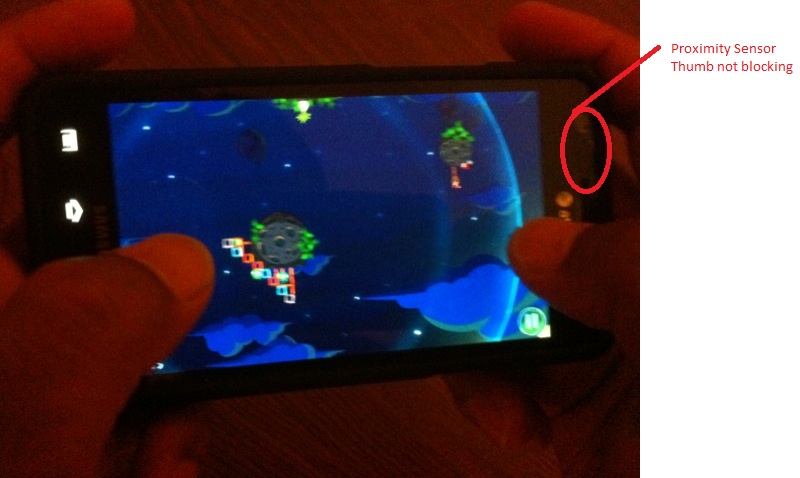}
\caption{Orientation 2 (O2): Thumb Not Blocking the Proximity Sensor}
\label{o2}
\end{minipage}
\end{figure}

To test this hypothesis, the experimenter played a number of games
in the ``landscape mode'', such as FIFA or Modern Combat. This test was
performed while the phone was held in two orientations: 
\begin{enumerate}

\item[\textbf{O1:}] thumb blocking the proximity sensor (Figure \ref{o1})
\item[\textbf{O2:}] thumb not blocking the proximity sensor (Figure \ref{o2})

\end{enumerate}

The games
were played for a duration of about 15 minutes on each of our test devices.
Interestingly, the phone did register a few unlocking events in the second
orientation. A total of 4 such events occurred out of a total of 61 changes in sensor readings,
resulting in an FPR of 6.5\%. However, no unlockings were recorded in the first
orientation. This could be attributed to the fact that when the thumb was
blocking the sensor, it did not induce sudden fluctuations to the reading. On
the other hand, when the thumb was below the sensor, while playing a game, the
gestures, such as firing a gun or moving an object, would have triggered the
sensor. 

We also conducted tests to simulate a nearby adversary who may (deliberately) bumps into a user, while the phone is in the pocket, in order to trigger a false unlocking. We did 10 trials of this test, and no unlocking was recorded, leading to a FPR of $0\%$.

The results of all of our tests are summarized as a confusion matrix depicted
in Table \ref{tab:prox-exp}.  These indicate that the tap-wave-rub gesture
based on proximity sensing is quite robust to accidental unlocking in practice.
The only scenario where the phone could accidentally get unlocked was during a
touchscreen game play activity when the phone was held in a particular
orientation.  However, even this scenario provides an important insight.
Namely, a rubbing gesture is likely to be exhibited while playing games in the
landscape mode under one of the orientations only, the one where the thumb does
not block the sensor. This implies that, depending upon the location of the
proximity sensor on a phone, one of the orientations could be disabled by the
phone's OS to prevent this accidental unlocking from taking place.





%% file: discussion.tex

Overall, our experiment results in previous section show that both phone
tapping recognition, and hand waving, finger tapping or rubbing recognition can
serve as two different means to infer the ``right'' human activity in order to
unlock the use of sensitive services on smartphones, thus preventing
unauthorized and stealthy access from malware.  Both result in quite low error
rates, FNR as well as FPR, demonstrating the effectiveness of our approach. We
believe that FNR can be further reduced as users become more and more familiar
with the underlying gestures.   

In this section, we compare the performance of the two gesture recognition
schemes and discuss issues relevant to gesture recognition for human activity
inference in general.

\subsection{Effect of Orientation}

For phone tapping recognition, we use a single three-axis accelerometer to infer
the force due to tapping by measuring accelerations on the three axes. When the
phone is held in the hand with different ways or under different orientations, the
accelerometer can be tilted around the three axes. This means that the same external force may
produce different accelerations along the three axes of the accelerometer.
Hence, successful recognition rate (or FNR) can be affected due to different
phone orientations. However, as long as the phone is held in the similar way,
thus the same orientation, when a user performs tapping (which is usually the
case), the tilting effect can be minimized. We emphasize that for all three tapping gestures, we
achieve similar high recognition rates (or low FNR) as other hand gestures in
previous study \cite{secret_handshakes,uwave}. On the other hand, orientation
or the tilt effect does help us achieve very low FPR as we explained in Section
\ref{sec:tap-exp}.

Our experiment results show that phone's orientation can affect the FPR of our
proximity sensor based scheme. This is because in a certain orientation, it
might be possible for hand movements to accidentally trigger the proximity
sensor. As argued in Section \ref{sec:wave-exp}, we suggest the OS can disable
such a ``dangerous'' orientation to prevent this accidental unlocking from
taking place. Luckily, due to the location of the proximity sensor (near the
ear piece), only one of the orientations need to be disabled, while the other
can safely be permitted, without lowering the overall usability. Another
possibility to avoid such an accidental unlocking in our scheme is to infer the
orientation of the device based on accelerometer data. If a dangerous
orientation is detected, the access can be disabled by default. 

\subsection{Implicit vs. Explicit Gestures}

As outlined in Section \ref{sec:overview}, gesture recognition can be implicit
or explicit. In implicit gesture recognition, a user's natural movement to
access a certain service is captured for human activity inference. The good
thing about implicit gesture recognition is that it does not change the usage
model of existing smartphone applications and transparently enforces access
control to sensitive services.  The phone tap gesture is an example of such an
implicit gesture required for accessing the NFC service. Not every service can
be associated with a particular gesture, however. 

In explicit gesture recognition, a user is required to perform a certain hand
movement to unlock a service.  The hand wave or finger tap/rub gesture is an
example of such an explicit gesture. As long as gesture recognition is robust
(with low FPR and FNR) and lightweight, multiple services can share the same
explicit gesture. This is indeed the case for our explicit gestures. Given that
our phone tap gesture performed quite well, it can also be used as an explicit
gesture to access a service besides NFC. In this case, the user would simply
need to tap her phone with her hand or a nearby object to unlock a service. 

\subsection{User Characteristics}

The phone tap gesture is user-dependent as it captures the user's own (perhaps
unique) feature of the tapping movement for recognition. A previous study shows
that there is significant variation among different participants even for the
same predefined gesture \cite{uwave}. Indeed, user-dependent gestures can be
and has been used for user authentication
\cite{gafurov2006biometric,conti2011mind}. As for phone tap, different users
may hold the phone in different ways and use different forces to perform the
gesture. This means that this gesture may provide user authentication, and
thus protection against unauthorized access to services in case the phone is
lost/stolen. Validation of this hypothesis required further investigation.
However, when a phone is shared by multiple users, say by different family
members, each has to register his/her own tap template in order to use our
approach. This training is only one-time and will not undermine the usability
of our approach. 

On the other hand, our hand wave, and finger tap/rubbing gestures is quite
user-independent as shown in our experiment results. That means, the service
can be shared by multiple users without registering his/her own template. In
fact, there is no template employed in this scheme. This scheme therefore may
offer a higher level of convenience to the users and easily adoptable. 
Of course, it does not prevent unauthorized use of such service
when the phone is stolen.

\subsection{Social Engineering Attacks}

The low FPR in our experiments demonstrates the low likelihood that an
application will gain access to critical resources without the knowledge of the
user. This is based on an assumption that a malware program can not emulate the
required gestures. However, it is possible for a competing malware writer to
fool a user by launching a social-engineering attack. For example, a malware
developer can design a game such that user has to move his phone in certain
ways mimicking the gesture of tapping, or a game in which a frequently pressed
button is placed very close to the proximity sensor, which may trigger wave or
rubbing gesture. 
Also, malicious program can wait for the legitimate program to ask for the gesture permission and while user permits the legitimate application, the eavesdropping malware can use the gesture along with that application to trigger malicious activity.
While such attacks are likely, they still require the malware
program to constantly wait for, and synchronize with, the desired user
gestures, which may make these programs easily detectable by the OS.
Nevertheless, our approach still significantly raises the bar against many existing 
malware attacks, a prominent advancement in
start-of-the-art in smartphone malware prevention.  


\subsection{Other Explicit Gestures}

To further extend our approach, it is natural to consider other explicit
gestures, or other ways to infer the gestures. So far, we detected hand waving
based on proximity sensor data. The ambient light sensor readings can also be
used in this context, as shown in another recent work \cite{light-wave}. The
advantage of this sensor is that the hand would not need to be very close to
the phone. This approach also requires accelerometer data to reduce false
positives, as simple phone movement may also trigger light values.

Accelerometer-based explicit gestures are definitely applicable for our purpose
and some of the previously developed gesture recognition schemes
\cite{secret_handshakes,uwave} might be suitable. However, such gestures may
not be as intuitive and lightweight as the explicit gesture proposed in this
paper and may require a higher level of user diligence. Besides accelerometers,
other sensors that come to mind, and may be useful for our case, are
magnetometers, and ambient temperature sensors. It does not seem possible,
however, to trigger a magnetometer without carrying a specialized equipment or
object, such as a magnet. Similarly, fluctuating the environmental temperature
via human actions seem rather difficult.

\subsection{Power Efficiency}

Another important aspect to cover in our work is the power consumed by the
sensors while trying to capture the user gesture. Since the battery-life is one
of the most important factors to be considered for user's day-to-day activity,
our design needs to be battery-friendly. 

Our approach is indeed quite power-efficient. The gestures proposed by our
design are short (last only a few seconds) and thus we only need to turn on the
underlying sensors momentarily.  Only when the user requests access to a
particular resource, the sensor is turned on and the (explicit) gesture detection
algorithm is executed.\footnote{For the sake of our experiments, we have turned
on the sensors all the time. This was done so that we can determine the FPRs
via our experiments.} Once the kernel captures the required gesture, the
permission will be granted to the active application and sensor will be turned
off. If kernel fails to capture the gesture within a certain duration, the
sensor will be turned off and application will be denied to use the resource.
In the case of NFC services or implicit gestures, tapping detection algorithm is executed only
when the NFC reader detects a nearby tag.
Moreover, our gesture detection algorithms
themselves are very lightweight and thus only require negligible amount of
power. 
